\documentclass[fleqn,twoside]{article}
\usepackage{espcrc2}
\newcommand{\Slash}[1]{\ooalign{\hfil/\hfil\crcr$#1$}}

\newcommand{\AmS}{{\protect\the\textfont2
  A\kern-.1667em\lower.5ex\hbox{M}\kern-.125emS}}

\title{Construction of Supersymmetric Nonlinear Sigma Models
on Noncompact Calabi-Yau Manifolds with Isometry}

\author{Kiyoshi Higashijima\address[OU]{Department of Physics,
Graduate School of Science, Osaka University, \\
 Toyonaka, Osaka 560-0043, Japan }%
    \thanks{Work supported in part by the Grant-in-Aid for
    Scientific Research},
    Tetsuji Kimura\addressmark[OU]
    and
    Muneto Nitta\address{Department of Physics, Purdue University,
    West Lafayette, IN 47907-1396, USA}\thanks{Work supported by the U.S.
Department of Energy under grant DE-FG02-91ER40681 (Task B)}}

\begin{document}

\begin{abstract}
We propose a class of ${\cal N}=2$ supersymmetric nonlinear sigma models
on the noncompact Ricci-flat K\"ahler manifolds, interpreted
as the complex line bundles over the hermitian symmetric spaces.
K\"ahler potentials and Ricci-flat metrics for these manifolds
with isometries are explicitly constructed by using the techniques
of supersymmetric gauge theories. Each of the metrics contains
a resolution parameter which controls the size of these base
manifolds, and the conical singularity appears when the parameter vanishes.
\vspace{1pc}
\end{abstract}

\maketitle

\section{Introduction}
Nonlinear sigma models (NL$\sigma$Ms) in two dimensions are interesting
for several reasons.
They help us to understand various non-perturbative phenomena in four
dimensional gauge theories such as confinement or dynamical mass
generation. They also provide the description of strings propagating
in the curved space-time. In the latter case, the consistency of strings
requires the conformal invariance of the NL$\sigma$Ms. In the case of
superstrings, furthermore, NL$\sigma$Ms have to possess the ${\cal N}=2$
world sheet superconformal symmetry. Although all two-dimensional
NL$\sigma$Ms with ${\cal N}=2$ supersymmetry have ${\cal N}=2$
superconformal symmetry at the tree level, they suffer from
superconformal anomaly at the quantum level. Since these anomalies are
proportional to the Ricci curvatures $R_{ab}$ of the background (target)
manifolds, ${\cal N}=2$ supersymmetric NL$\sigma$Ms taking values on
the target manifolds with vanishing Ricci curvatures, provide the
possible candidates of the consistent superstrings propagating in
the curved space-time. They are also candidates of finite quantum
field theories in two dimensions. ${\cal N}=2$ supersymmetry
in two dimensions requires the target manifolds of NL$\sigma$Ms are
K\"{a}hler manifolds. The K\"{a}hler manifolds with vanishing
Ricci curvatures are called the Calabi-Yau manifolds. Therefore,
it is important to study the ${\cal N}=2$ supersymmetric NL$\sigma$Ms
defined on the Calabi-Yau manifolds.

To find the metric of the Ricci-flat manifolds, we have to solve
the Einstein equation $R_{ab}=0$ for the vacuum. If the manifold has
enough symmetries, we can reduce the partial differential equation
to an ordinary differential equation, which is usually easy to solve.
In compact Calabi-Yau manifolds, however, there is no isometry and not
a single explicit metric is known in this case. In some noncompact
Calabi-Yau manifolds, the number of isometries is sufficient to reduce
the Einstein equation to an ordinary differential equation.
In this article, we report our recent progress to obtain
the explicit metric for certain class of the noncompact Calabi-Yau
manifolds \cite{HKN1,HKN2,HKN3}.

\section{NL$\sigma$Ms with ${\cal N}=2$ Supersymmetry in Two Dimensions}
${\cal N}=2$ supersymmetry requires that both scalars
$A^a(x)$ and fermions $\psi^a(x)$ are complex fields.
The scalar fields $A^a(x) \quad (a=1,\cdots,N)$ are coordinates
of the target manifold, whose metric is given
by the K\"{a}hler potential $K(A,{\bar A})$ which characterizes the theory:
\begin{equation}
 g_{a{\bar b}}(A,{\bar A})
  = {\partial^2 K(A,{\bar A}) \over \partial A^a \partial {\bar A}^{\bar b}}
  = K,_{a{\bar b}} (A,{\bar A}).\label{kmetric}
\end{equation}
The lagrangian for ${\cal N}=2$ supersymmetric NL$\sigma$M is given by
\begin{eqnarray}
{\cal L}(x) =&& \ g_{a \bar{b}} \partial_{\mu} A^a \partial^{\mu}
{\bar A}^{\bar{b}} + i g_{a{\bar b}} \bar{\psi}^{\bar b} ( \Slash{D} \psi)^a
\nonumber\\
&&+ \frac{1}{4} R_{a {\bar b}c {\bar d}} \psi^a \psi^c
\bar{\psi}^{\bar b} \bar{\psi}^{\bar d} ,\label{lagrangian}
\end{eqnarray}
where the covariant derivative and the Riemann curvature are defined by
\begin{eqnarray}
(D_{\mu}
\psi)^a &=& \partial_{\mu} \psi^a + \partial_{\mu} A^b \Gamma^a{}_{b c}
\psi^c\\
R_{a {\bar b}c {\bar d}}&=&\partial_c\partial_{\bar d}g_{a{\bar b}}
-g^{e{\bar f}}\partial_cg_{a{\bar f}}\partial_{\bar d}g_{e{\bar b}},
\end{eqnarray}
with the connection being
\begin{equation}
 {\Gamma^c}_{ab} = g^{c{\bar d}} g_{b{\bar d}, a}
                 = g^{c{\bar d}} K,_{\,ab{\bar d}}.
\end{equation}
Since all geometrical quantities are derived from
the K\"{a}hler potential $K$,
all we have to do is to assume an appropriate symmetry of the scalar
function $K$. Symmetries of a scalar function is far simpler
than the symmetries of the tensor $g_{a{\bar b}}$.

\section{Noncompact Calabi-Yau Manifolds}

Although the NL$\sigma$M defined by (\ref{lagrangian}) is conformally
invariant at the tree level, it suffers from the conformal anomaly
proportional to the Ricci curvature at the quantum level\cite{AG}.
To obtain the conformally invariant theory at the quantum level,
we impose the Ricci-flatness condition

\begin{equation}
R_{a{\bar b}}=0\label{ricciflat}
\end{equation}
to the target manifold.

 In order to reduce an Einstein equation to an ordinary differential
 equation of single variable, we assume our target manifold $M_1$ has
 a symmetry, which carries each point of the manifold to a certain
 point on a line, which is parameterized by a single variable.
 In other word, any point of our manifold $M_1$ can be reached from
 a point on a line by a suitable action of some symmetry group $G$.
 If any point of the manifold can be reached from a single point,
 say the origin, by appropriate action of a symmetry group $G$,
 the manifold is a homogeneous coset space $G/H$ where $H$ denotes
 a subgroup of $G$. The local structure of our manifold is, therefore,
 the product of a homogeneous space $M=G/H$ and a line.
 Since ${\cal N}=2$ supersymmetry requires the manifold $M_1$ to be
 a K\"{a}hler manifold, we assume that the homogeneous space $M=G/H$
 is {\it K\"{a}hler} by itself and the line is a {\it complex} line.
 Furthermore, we assume the K\"{a}hler coset space $M=G/H$ is
 a compact {\it Einstein manifold}, which satisfies the Einstein
 equation with a positive cosmological constant $h>0$
\begin{equation}
R_{m{\bar n}} \ = \ h g_{m{\bar n}} , \label{Einstein}
\end{equation}
where we use the indices $m,n$ for the compact K\"{a}hler manifold
$M=G/H$ whose complex dimension is $d=dim(G/H)/2$. The complex
dimension of the Ricci-flat manifold $M_1$ is $D=1+d$.

Our total space $M_1$ is parameterized by the complex coordinates
$\phi_m \ (m=1,\cdots,d)$ and a complex number $\rho$ that represents
the complex line. We denote the K\"{ahler} potential of the compact
K\"{a}hler-Einstein manifold $M$ by $\Psi(\phi, \bar{\phi})$.
Our assumption for the K\"{a}hler potential of the Ricci-flat
manifold $M_1$ is as follows:
\begin{equation}
K(A,{\bar A})=K(X),
\end{equation}
where
\begin{equation}
X=\log|\rho|^2+h\Psi(\phi, \bar{\phi}).\label{defx}
\end{equation}
Now, if we notice
\begin{equation}
R_{a{\bar b}}=-g^{c{\bar d}}R_{a{\bar b}c{\bar d}}
=-\partial_a\partial_{\bar b}\log{\det{(g_{c{\bar d}})}} \label{ricci}
\end{equation}
the Ricci-flat condition (\ref{ricciflat}) takes a very simple form
\begin{equation}
\det{(g_{c{\bar d}})}= |{\rm hol.}|^2,\label{flatdet}
\end{equation}
where $\rm hol.$ is a holomorphic function of the coordinates
$\phi$ and $\rho$.

Components of the metric for the whole space $M_1$
\begin{equation}
 g_{a{\bar b}} \ = \ \left(
 \begin{array}{cc}
 g_{\rho \bar{\rho}} & g_{\rho {\bar n}} \\
 g_{m {\bar{\rho}}} & g_{m{\bar n}}
 \end{array} \right) \; ,
\end{equation}
are given by
\begin{eqnarray}
 g_{\rho \bar{\rho}}
\ &=& \
K''{ \partial X\over\partial \rho} {\partial X\over \partial
{\bar \rho}} \; ,
\label{grr} \\
g_{\rho {\bar n}}
\ &=& \
K'' {\partial X\over \partial \rho} {\partial X\over \partial
{\bar \phi}^{\bar n}} \; , \label{grn}\\
g_{m{\bar n}}
\ &=& \
K'' {\partial X\over \partial \phi^m} {\partial X\over \partial
{\bar \phi}^{\bar n}}
+ K' \frac{\partial^2 X}{\partial \phi^m \partial
{\bar \phi}^{\bar n}} \; ,\label{gmn}
\end{eqnarray}
where the prime denotes the differentiation with respect to
the argument $X$.
Using equations
\begin{equation}
{\partial X / \partial \rho} = 1/ \rho (\rho \neq 0),\quad
\frac{\partial^2 X}{\partial \phi^m \partial {\bar \phi}^{\bar n}}
=  hg_{m{\bar n}},
\end{equation}
we can express the determinant of the metric of $M_1$ in terms of
the determinant of the metric of $M=G/H$
\begin{eqnarray}
 \det{ g_{a {\bar b}}}
 \ &=& \ g_{\rho {\bar \rho}} \cdot
   \det (g_{m{\bar n}} - g_{\rho{\bar \rho}}^{-1} g_{m\bar{\rho}}
   g_{\rho{\bar n}})\\
 \ &=& \
 \frac{1}{|\rho|^2} K''
 (K')^d \cdot  \det (h g_{m{\bar n}}) \; .
 \label{metric-determinant}
\end{eqnarray}

Since we have assumed that the K\"{a}hler coset space is
an Einstein space, we can easily calculate the determinant
of its metric.
By substituting the expressions of the Ricci curvature (\ref{ricci})
and the metric $g_{m{\bar n}}$ in terms the K\"{a}hler potential
$\Psi$ (\ref{kmetric}) to the Einstein equation (\ref{Einstein}),
we obtain
\begin{equation}
R_{m{\bar n}}=-\partial_m\partial_{\bar n}\log{\det{(g_{k{\bar l}})}}
=h{\partial^2 \Psi(\phi,{\bar \phi}) \over \partial \phi^m
\partial {\bar \phi}^{\bar n}},
\end{equation}
which can be integrated to give a convenient formula for
the determinant of the metric in K\"{a}hler-Einstein manifold $M$
\begin{equation}
\det (g_{m{\bar n}}) \ = \ e^{ - h \Psi} |{\rm hol.}|^2.\label{detformula}
\end{equation}

Combining the formula (\ref{metric-determinant}),
(\ref{detformula}) and (\ref{flatdet}), we obtain the Ricci-flatness
condition for the K\"{a}hler potential of the manifold $M_1$
\[
 \det (g_{a {\bar b}})
 \ = \
  e^{- X} K'' ( K')^{d} |{\rm hol.}|^2 =(const.)\cdot |{\rm hol.}|^2 ,
\]
from which we find an ordinary differential equation for
the K\"{a}hler potential:
\begin{equation}
e^{-X}  {d\over dX}( K')^D \ = \ a  , \label{ODE}
\end{equation}
where $a$ is a constant.

The equation (\ref{ODE}) is easily solved to yield
\begin{equation}
 K' \ = \ \big( \lambda e^X + b \big)^{\frac{1}{D}} \; ,
\label{sol}
\end{equation}
where $\lambda$ is a constant related to $a$ and $D$,
and $b$ is an integration constant, interpreted as a resolution
parameter.
To calculate the metric (\ref{grr})(\ref{grn})(\ref{gmn}) of
the Ricci-flat manifold $M_1$, the K\"{a}hler potential itself
is not necessary, though it is easily obtained by integrating (\ref{sol}).

Our method can be applied to any K\"{a}hler-Einstein coset manifold $M$,
however, we need the K\"{a}hler potential $\Psi$ of $M$ to obtain
the explicit expression for the metric of the noncompact Calabi-Yau
manifold $M_1$. Fortunately, the
general prescription to calculate $\Psi$ is given by Ito, Kugo and
Kunitomo\cite{IKK}. In this respect, our results gives the explicit
realization of the method of Page and Pope, who constructed
Ricci-flat manifolds by introducing a complex line as a fibre
over a K\"{a}hler-Einstein manifold\cite{PP}.

Here, we further require that the K\"{a}hler coset space $M$ is
a symmetric space, called the hermitian symmetric space, then
the K\"{a}hler potential can be explicitly obtained by the gauge
theory technique\cite{HKN2,HN}. Therefore, we find a new class
of Ricci-flat manifolds corresponding to various hermitian
symmetric spaces with $h=C_2(G)/2$. For $M=CP^{N-1}$, $M_1$
reduces to the manifold obtained by Calabi\cite{Ca}.

\end{document}